\documentclass[aip,twocolumn,letterpaper]{revtex4}

\usepackage{fullpage}

\usepackage{graphics}
\usepackage{epsfig}

\usepackage{xcolor}

\usepackage[margin=1.0in]{geometry}

\begin{document}
\title{Example of exponentially enhanced magnetic reconnection driven by a spatially-bounded and laminar ideal flow }
\author{Allen H Boozer and Todd Elder}
\affiliation{Columbia University, New York, NY  10027\\ ahb17@columbia.edu}

\begin{abstract}

In laboratory and natural plasmas of practical interest, the spatial scale $\Delta_d$ at which magnetic field lines lose distinguishability differs enormously from the scale $a$ of magnetic reconnection across the field lines.  In the solar corona, plasma resistivity gives $a/\Delta_d\sim10^{12}$, which is the magnetic Reynold number $R_m$.  The traditional resolution of the paradox of disparate scales is for the current density $j$ associated with the reconnecting field $B_{rec}$ to be concentrated by a factor of $R_m$ by the ideal evolution, so $ j\sim B_{rec}/\mu_0\Delta_d$.  A second resolution is for the ideal evolution to increase the ratio of the maximum to minimum separation between pairs of arbitrarily chosen magnetic field lines, $\Delta_{max}/\Delta_{min}$, when calculated at various points in time.  Reconnection becomes inevitable where $\Delta_{max}/\Delta_{min}\sim R_m$.  A simple model of the solar corona will be used for a numerical illustration that the natural rate of increase in time is linear for the current density but exponential for $\Delta_{max}/\Delta_{min}$.  Reconnection occurs on a time scale and with a current density enhanced by only $\ln(a/\Delta_d)$ from the ideal evolution time and from the current density $B_{rec}/\mu_0a$.  In both resolutions, once a sufficiently wide region, $\Delta_r$, has undergone reconnection, the magnetic field loses static force balance and evolves on an Alfv\'enic time scale.  The Alfv\'enic evolution is intrinsically ideal but expands the region in which $\Delta_{max}/\Delta_{min}$ is large.

\end{abstract}

\date{\today} 
\maketitle

\section{Introduction}

Magnetic reconnection has an interesting early history \cite{Cargill:2015} and was defined in 1956 by Parker and Krook \cite{Parker-Krook:1956} as the ``\emph{severing and reconnection of lines of force.}"

The ideal evolution of magnetic fields is the opposing concept to magnetic reconnection.   Newcomb showed in 1958 that magnetic field lines move with a velocity $\vec{u}_\bot$ and do not break, if and only if the field obeys the ideal evolution equation \cite{Newcomb} 
\begin{equation}
\frac{\partial \vec{B}}{\partial t}=\vec{\nabla}\times(\vec{u}_\bot\times\vec{B}). \label{Ideal ev}
\end{equation}

The exact equation for the evolution of $\vec{B}$ is Faraday's Law, $\partial \vec{B}/\partial t=-\vec{\nabla}\times\vec{E}$.  The ideal evolution of a magnetic field that is embedded in a plasma is broken \cite{Boozer:ideal-ev} when the electric field parallel to the magnetic field, $\vec{E}_{||}$, cannot be balanced by the parallel component of the gradient of a well-behaved potential, $\vec{\nabla}\Phi$.   The components $\vec{E}_\bot+\vec{\nabla}_\bot\Phi$, which are perpendicular to $\vec{B}$, are associated with the velocity $\vec{u}_\bot$ of the magnetic field lines, $\vec{u}_\bot =(\vec{E}_\bot+\vec{\nabla}_\bot\Phi)\times\vec{B}/B^2$.   

Two effects that cause deviations from an ideal evolution are resistivity along the magnetic field $\eta$, which contributes to $E_{||}$ as $\eta \vec{j}_{||}$, and electron inertia, which contributes \cite{Spitzer} to $E_{||}$ as $(c/\omega_{pe})^2 \partial j_{||}/\partial t$.  The electron skin depth, $c/\omega_{pe}$, is the speed of light divided by the electron plasma frequency.   

The Oxford English Dictionary defines a paradox as ``\emph{a strongly counter-intuitive} (statement)\emph{, which investigation...may nevertheless prove to be well-founded or true.}"  Both the resistivity and the electron skin depth are so small in many plasmas of practical interest that it is paradoxical that magnetic reconnection could be of any relevance.  Yet it is.  Approximately 1500 papers have been written on magnetic reconnection, which demonstrate the importance of the topic to the understanding of both laboratory and naturally occurring plasmas.  The speed and prevalence of magnetic reconnection are so great that they must be derivable from the properties of Equation (\ref{Ideal ev}) for the ideal evolution of a magnetic field \cite{Boozer:ideal-ev}.

To define the reconnection paradox, let $a$ be a characteristic spatial scale over which $u_\bot$ varies across the magnetic field lines; the time scale for an ideal evolution is 
\begin{equation} \tau_{ev}\equiv \frac{a}{u_\bot}, \end{equation}
Resistivity spatially interdiffuses magnetic field lines, which implies that lines that approach each other closer than the distance $\Delta_d =\sqrt{(\eta/\mu_0)\tau_\eta}$ remain distinguishable only for a time $t$ less than $\tau_\eta$.  The ratio of the resistive time scale of the overall system to the evolution time scale $\tau_{ev}$ is the  magnetic Reynolds number 
\begin{equation}
R_m \equiv \left(\frac{a^2}{\eta/\mu_0}\right) \left(\frac{u_\bot}{a}\right)=\frac{\mu_0 u_\bot}{\eta}a.
\end{equation} 
Magnetic field lines that come closer than $\Delta_d=a/R_m$ lose their distinguishability, where $R_m\sim 10^{12}$ in the solar corona.  

The electron inertia causes evolving magnetic field lines that approach each other closer than $\Delta_d=c/\omega_{pe}$, the electron skin depth, to become indistinguishable, Appendix C of \cite{Boozer:null-X}. In the solar corona $a/(c/\omega_{pe})\sim 10^9$.

What is paradoxical is that magnetic field lines are observed to reconnect and become indistinguishable over a region of width $a$ across the magnetic field lines on a time scale only an order of magnitude or so longer than the characteristic ideal-evolution time of the magnetic field, $\tau_{ev}$.  Effects such as the resistivity would be expected to cause a loss of distinguishability of magnetic field lines in the solar corona only on a time scale of order $10^{12}$ times longer than $\tau_{ev}$.  This paradox was clearly described in 1988 by Schindler, Hesse, and Birn \cite{Schindler:1988}. 

Schindler et al \cite{Schindler:1988} sought to resolve the paradox of reconnection occurring over a region of far greater width than $\Delta_d$.  In their resolution, the current density would become and remain extremely large, $j\sim B_{rec}/(\mu_0\Delta_d)$, in a layer of width $\Delta_d$, where $B_{rec}$ is the part of the magnetic field that is reconnecting.  This assumption has formed the basis of most of the reconnection literature from even before Schindler et al developed their reconnection theory.  Nonetheless, it is difficult to understand how an ideal evolution would result in such a strong current.  Most of the reconnection literature has not dealt with that issue but has focused instead on how such a large current density could be maintained if it were initially present.  A foundational publication for recent work on the maintenance issue, with more than two hundred citations, is the 2010 paper by Uzdenksy, Louriero, and Schekochihin \cite{Loureiro:2010} on plasmoids.

The paradoxical speed of magnetic reconnection when non-ideal effects are extremely small has analogues with other phenomena, such as thermal equilibration in air.  The analogy between magnetic reconnection and thermal equilibration in a room was demonstrated \cite{Boozer:rec-phys} by Boozer in 2021.  In both reconnection and thermal equilibration, the time scale of the relaxation of the ideal constraint is the ideal evolution time $\tau_{ev}$ times a term that is logarithmically dependent on the ratio of the non-ideal time scale divided by the $\tau_{ev}$.  This explains why a radiator can heat a room in tens of minutes rather than the several weeks that would be expected from thermal diffusion alone.  Many may find the subtle mathematics used in \cite{Boozer:rec-phys} difficult to follow.  Here a simple model of magnetic field evolution driven by footpoint motion, as in the solar corona, is used to illustrate the general principles that were explained in the companion paper  \cite{Boozer:rec-phys}.

The equation for temperature equilibration in air moving with a divergence-free velocity $\vec{v}(\vec{x},t)$ is $\partial T/\partial t+\vec{v}\cdot \vec{\nabla}T=\vec{\nabla}\cdot (D \vec{\nabla}T)$, which is the standard form for the advection-diffusion equation.  In many problems of practical importance,  the P\'eclet number, $va/D$, is many orders of magnitude greater than unity.  The diffusion coefficient for magnetic field lines is $\eta/\mu_0$, so the magnetic Reynolds number would be better called the magnetic P\'eclet number.   

The classic paper on the advection-diffusion equation was written in 1984 by Hassan Aref \cite{Aref;1984}  and has had over two thousand citations.  This paper showed that a laminar flow can equilibrate $T$ on the evolution time $a/v$ times a logarithm of the P\'eclet number as $va/D\rightarrow\infty$.  Before Aref's paper, it had been assumed a turbulent $\vec{v}$ was required to obtain fast equilibration.  Section  I.B in an article \cite{Aref:2017} published in the Reviews of Modern Physics discusses the merits of the use of turbulent versus laminar flows to speed the mixing of fluids.  What that article did not discuss is that for a given maximum flow speed $v$ a laminar flow can give a faster mixing over a large region than a turbulent flow.  A demonstration starts with the last sentence on p. 3 of \cite{Boozer:rec-phys}.   The laminar part of a flow, which is defined as an average over the small spatial scales of $\vec{v}(\vec{x},t)$, generally dominates the advective part of the advection-diffusion equation.   A flow $\vec{v}(\vec{x},t)$, no matter how complicated and rapid the dependence on $\vec{x}$ and $t$, cannot directly produce  diffusive mixing but can exponentially enhance the speed with which diffusion can act over the spatial scale covered by the advection of the fluid.

The essential element in an enhanced relaxation by a laminar flow is that the flow be chaotic.  Using standard terminology, a flow is deterministic but chaotic when neighboring streamlines have a separation that increases exponentially with time.  Articles on the mathematics of deterministic chaos and topological mixing can easily be found on the web, but their importance to this paper is only that such effects are common.  As noted in \cite{Boozer:rec-phys}, although the condition that the flow be chaotic may sound restrictive, it is a non-chaotic natural flow that is essentially impossible to realize.  No special effort is required to achieve enhanced mixing by stirring.  Every cook knows that stirring enhances the mixing of fluids---no particular pattern of stirring or detailed computations are required.  


The terms chaotic and stochastic are sometimes considered synonyms in descriptions of flows or magnetic fields.  Here a distinction is made that is consistent with a distinction made in the mathematical literature.  Chaos, or more precisely deterministic chaos, when applied to a flow within a bounded region of space, means the streamlines of the flow are predictable but exponentiate apart throughout a finite fraction of that space, the region in which the flow is chaotic.  In contrast, a stochastic motion is not deterministic and has a random component on all time and spatial scales.  Stochastic flows were used by L. F. Richardson in a paper \cite{Richardson:1926} published in 1926 to explain the enhancement of diffusion produced by atmospheric flows.  His assumption was that that the velocity of air $\Delta\vec{x}/\Delta t$ has no well defined limit as $\Delta t$ goes to zero.  Weak magnetic field line stochasticity makes field lines indeterminant in a way that is similar to the the effect that Richardson's assumption has on streamlines of air.  The enhancement of reconnection by weak stochasticity of the magnetic field lines  was studied by A. Lazarian and E. T. Vishniac \cite{Lazarian:1999}  in  1999.   Here we consider only deterministic flows, which means velocities are well defined.  In an ideal magnetic evolution, Equation (\ref{Ideal ev}), deterministic flows give deterministic magnetic field lines. 
   

The evolution equation for a magnetic field is also of the advection-diffusion type but more subtly so than the equation for thermal equilibration.  The companion paper \cite{Boozer:rec-phys} derives two salient differences:  (1) A flow-enhanced equilbration of the temperature requires a dependence of $\vec{v}$ on at least two spatial coordinates to be energetically feasible, but flow-enhanced magnetic reconnection requires a dependence of $\vec{u}_\bot$ on all three spatial coordinates for energetic feasibility.  (2) In thermal relaxation, $\big|\vec{\nabla}T\big|$ increases exponentially with time but what one might think would be the analogous quantity for the magnetic field, $\vec{j}=\vec{\nabla}\times\vec{B}/\mu_0$ does not.  In the model developed in this paper, the increase in the current density is limited to being proportional to time.  Related limits on the current density are known for resonant perturbations to toroidal plasmas \cite{Hahm-Kulsrud} and for currents flowing near magnetic field lines that intercept a magnetic null \cite{Elder-Boozer}.

When the ideal flow velocity of a magnetic field $\vec{u}_\bot$ is chaotic, the ratio $\Delta_{max}/ \Delta_{min}$, the maximum to the minimum separation between two field lines at any particular time, tends to increase exponentially during the evolution when the closest approach of the lines is small, $\Delta_{min}/a\rightarrow0$.  Large scale reconnection occurs for magnetic field lines that satisfy $\Delta_{min}\lesssim\Delta_d$ and $\Delta_{max} \approx a$.  As will be seen, a simple laminar flow can be followed as $\Delta_{max}/ \Delta_{min}$ increases by more than nine orders of magnitude.

The model that will be used to illustrate features of magnetic reconnection is developed in Section \ref{sec:model}.  In this model, the magnetic field evolution is driven in a bounded plasma in the same way as footpoint motion drives the magnetic field in the solar corona.  

Section \ref{sec:evolution} will show that when the footpoint velocity is sufficiently small compared to the Alfv\'en speed in the plasma that important features can be determined without solving the difficult problem of determining $\vec{B}(\vec{x},t)$ throughout the plasma. These features include the exponentiation in time of the separation between neighboring magnetic field lines and the current density $j_{||}$ along each magnetic field line, or more precisely $K\equiv\mu_0j_{||}/B$.   A bound on the magnitude of $K$ is obtained, which increases linearly in time.  For these features, only the determination of the streamlines of a time-dependent flow in two dimensions is required.  

The simplicity of the treatment given in Section \ref{sec:evolution} depends on the evolution being slow compared to the Alfv\'en transit time.  This would be violated if the plasma became kink unstable.  Section \ref{sec:kink} shows that the flow example introduced in Section \ref{sec:model} ensures kink stability.   

Section \ref{sec:flux}  shows that the rate of production of the magnetic flux associated with a particular magnetic field line by the footpoint motion is the magnetic Reynolds number $R_m$ times larger than its rate of destruction by resistivity.  The sign of the flux differs from field line to field line in kink-stable evolutions, and the rate of flux production and destruction can be brought into balance by reconnection. 

Section \ref{sec:helicity} shows that even if the plasma becomes kink unstable, the magnetic helicity contained in the bounded plasma will increase without limit when $R_m>>1$ unless the footpoint motion obeys a constraint.  If the plasma were unbounded, an ever increasing helicity would presumptively result in a plasma eruption.   

Section \ref{sec:power} derives the power that must be supplied to maintain the specified footpoint motion.  For a steady-state flow, the required power increases linearly with $K$.  Section \ref{sec:discussion} discusses the results and their implications.


\section{Model of the solar corona \label{sec:model}}

Features of magnetic reconnection will be illustrated using a simplified model of magnetic loops in the solar corona.  The evolution of coronal loops is driven by the motion of their foopoints by photospheric flows.  

The description of the model has three parts: (1) Section \ref{sec:def cyl region} defines a spatially bounded perfectly-conducting cylinder that encloses the loop as well as the flow $\vec{v}_t$ in the top surface of the cylinder, which represents the photospheric motion.  (2)  Section \ref{sec:choice of v_t} explains the specific form chosen for $\vec{v}_t$. (3) Section \ref{sec:quantification of chaos} explains how the chaotic properties of $\vec{v}_t$ are quantified.

The numerical results for the streamlines of the flow $\vec{v}_t$  are given in Section \ref{sec:v-results}.

\begin{figure}
\centerline{ \includegraphics[width=2.0in]{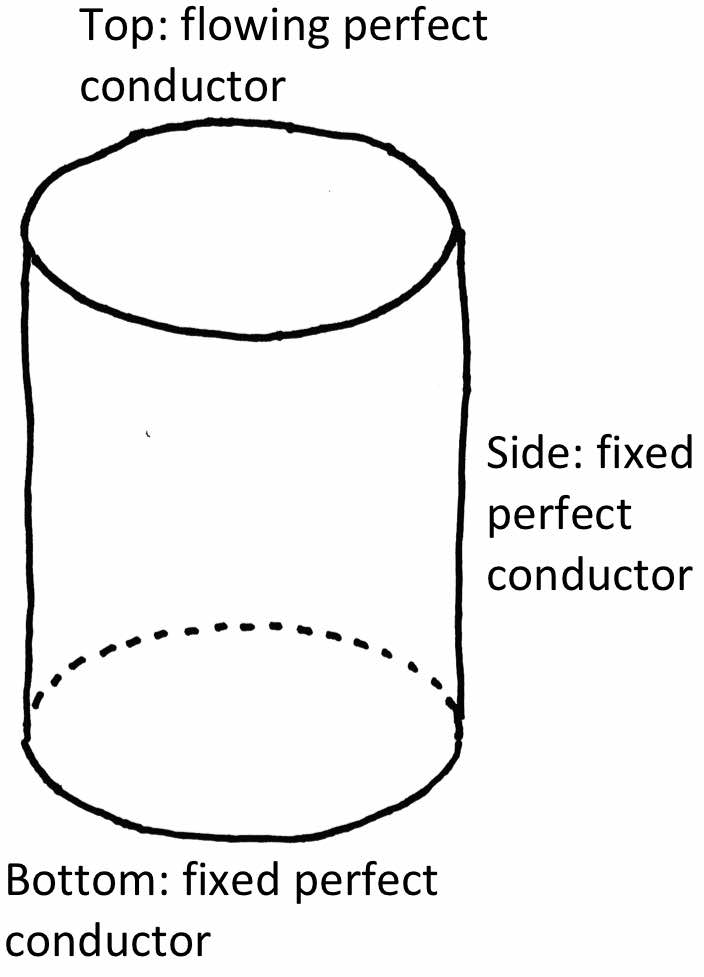}}
\caption{A perfectly conducting cylinder of  radius $a$ and height $L$ in $z$ encloses an ideal pressureless plasma.  All the sides of the cylinder are fixed except the top, which flows with a specified velocity $\vec{v}_t$.  Initially, $\vec{B}=B_0\hat{z}$. }  
\label{fig:cylinder}
\end{figure}

\subsection{Definition of cylindrical region \label{sec:def cyl region}}

The model that will be used to illustrate important features of magnetic reconnection consists of an ideal pressureless plasma enclosed by a perfectly conducting cylinder of radius $a$ and height $L$, Figure \ref{fig:cylinder}.  An ideal plasma means Equation (\ref{Ideal ev}) for an ideal magnetic evolution holds exactly.  The conducting surfaces of the cylinder are rigid except for a flowing top-surface, which provides the footpoint motion that drives the evolution.  

The flow of the top surface of the cylinder must be divergence free to avoid compressing the initial magnetic field $\vec{B}_0=B_0\hat{z}$ with $B_0$ a constant.   The implication is the flow must have the form
\begin{equation}
\vec{v}_t=\hat{z}\times \vec{\nabla}h_t(x,y,t), \label{v_t}
\end{equation}
where the stream function $h_t$ is chosen to represent a photospheric-like motion at one interception of a coronal magnetic loop.  For simplicity, the flow at the other photospheric interception is taken to be zero.  The effect of the flow is to produce a magnetic field orthogonal to $\vec{B}_0$. 

 The model illustrated in Figure \ref{fig:cylinder} is closely related to the well-known Parker Problem \cite{Parker:problem}, which was recently reviewed by  Pontin and Hornig \cite{Review:ParkerProb}.  It is also a simplified version of reconnection models of the corona published by Boozer \cite{Boozer:prevalence} and independently by Reid et al \cite{ Reid:2018} in 2018.  Although these two models are similar, the mathematical cause for reconnection emphasized in these papers is different.  Reid et al \cite{ Reid:2018} used an anomalous resistivity to ensure ``\emph{that resistivity, as opposed to a numerical diffusion, is responsible for any magnetic reconnection.}"  Boozer's paper and his more recent work \cite{Boozer:ideal-ev,Boozer:part.acc,Boozer:null-X,Boozer:rec-phys} emphasized that an imposed chaotic flow $\vec{v}_t$ would make magnetic reconnection exponentially sensitive to any departures from an ideal evolution---even effects due to numerics rather than physics.


\subsection{Choice of $\vec{v}_t$ \label{sec:choice of v_t}}

The divergence-free velocity $\vec{v}_t$ is specified by the stream function $h_t(x,y,t)$, Equation (\ref{v_t}).   Using Cartesian coordinates, the stream function is the Hamiltonian for the streamlines of the flow on the top of the cylinder $dx/dt= -\partial h_t/\partial y$ and $dy/dt = \partial h_t/\partial x$.  

The flow should satisfy certain conditions, and these are easier to impose in cylindrical coordinates $r,\theta,z$ where $x=r\cos\theta$ and $y=r\sin\theta$.  The vertical part of the cylindrical shell is at $r=a$.  The steam function $h_t$ should be chosen so $h_t$, $dr/dt$, $d\theta/dt$, and the radial gradient of $d\theta/dt$ are all zero at $r=a$.  This ensures that the streamlines can never strike the $r=a$ boundary of the cylinder, $\vec{v}_t=0$ at $r=a$, and extremely large currents do not form in the plasma near $r=a$.  

A form for $h_t$ that satisfies these conditions is
\begin{eqnarray}
h_t(r,\theta,t)&=& \tilde{h}(x,y,t) \left(1-\frac{r^2}{a^2}\right)^3 e^{-\lambda^2 r^2/a^2},\label{h}
\end{eqnarray}
where $\lambda^2$ is a constant.  A large $\lambda^2$ restricts the evolution-driven region to be far from the confining cylindrical walls.  In the studies reported here, $\lambda^2=0$.  The stream function $h_t$ is fully specified when $\tilde{h}(x,y,t)$ is given.  

Flows that carry footpoints over a scale comparable to $a$ are the most effective at producing a rapid reconnection \cite{Boozer:rec-phys}.  These are slowly varying terms in $x$ and $y$, which in simple forms are
\begin{eqnarray}
\tilde{h} &=&\frac{a^2}{\tau}\Big[ c_0 \cos\left( \omega_0 \frac{t}{\tau}\right) + c_1 \frac{x}{a}\cos\left( \omega_1 \frac{t}{\tau}\right) \nonumber\\
&&\hspace{0.01in} +  c_2 \frac{y}{a} \sin\left( \omega_2 \frac{t}{\tau}\right)\Big\} + c_3 \frac{xy}{a^2}\cos\left( \omega_3 \frac{t}{\tau}\right) \Big], \hspace{0.2in}  \label{h-tilde}
\end{eqnarray}
where $\omega_0$, $\omega_2$, and $\omega_3$ are three frequencies, which are generally incommensurate, and $c_0$, $c_1$, $c_2$, and $c_3$ are dimensionless amplitudes.  The amplitudes and frequencies used in the calculations are $c_0=0$, $c_1=c_2=c_3=1/4$, $\omega_1 = 6\pi$, $\omega_2 = 4\pi$, and $\omega_3 = 0$. 

The $c_0$ term by itself would produce streamlines that lie on circles, which produce a large parallel current density with a long correlation distance.  Such current densities tend to be ideal-kink unstable, Section \ref{sec:kink}, which would violate the assumption of a slow evolution compared to Alfv\'enic. As will be shown in Section \ref{sec:evolution} the analysis is greatly simplified when the evolution is slow compared to Alfv\'enic.  Consequently, the coefficient $c_0$ was chosen to be zero. A shown by Reid et al \cite{Reid:2020}, the kinks produced by pure circular motions lead to a large scale exponential separation of neighboring magnetic field lines.  The only reason for choosing $c_0=0$ is to be able to study a rigorously correct but simple example.

For determining the chaotic region associated with a particular initial condition $x_0,y_0$, it is advantageous for the frequencies to be commensurate because then a Poincar\'e plot can be constructed using the time-periodic points. When the frequencies are commensurate, $\tau$ is the periodicity or transit time of $\tilde{h}$.    The exponentiation, which as explained in Section \ref{sec:quantification of chaos}  is measured by the Frobenius norm, can be calculated whether the frequencies are commensurate or not.

The actual value of $\tau$ is arbitrary.  It is effectively the unit of time.  Similarly the distance $L$ can be chosen arbitrarily.  Only dimensionless ratios are important. The Alfv\'en speed $V_A$ only enters through the constraint, $L/\tau<<V_A$.

A circular conducting-cylinder is easier to discuss and does not complicate the computations of this paper, but a cylinder with a square cross section simplifies more complete simulations.  In a square cylinder, the factors of $1-r^2/a^2$ in Equation (\ref{h}) are replaced by $(1-x^2/a^2)(1-y^2/a^2)$.


\subsection{Quantification of chaos \label{sec:quantification of chaos} }

The stream function $h_t$ is the Hamiltonian for the stream lines of the flow $\vec{v}_t$ with $(x,y)$ the canonical variables.  The most important question about a specific flow is whether its streamlines are chaotic.  

Neighboring chaotic streamlines have a separation $\vec{\delta}$ that depends exponentially on time.  Neighboring means separated by an infinitesimal distance.  The exponentiation of neighboring streamlines can be defined streamline by streamline.  The $t=0$ location of a streamline $(x_0,y_0)$ defines that streamline, which at time $t$ has the position $\vec{x} =x(x_0,y_0,t) \hat{x}+ y(x_0,y_0,t) \hat{y}$, where $(\partial \vec{x}/\partial t)_{x_0y_0}\equiv d\vec{x}/dt=\vec{v}_t$.  

To determine not only the streamline that is at $(x_0,y_0)$ at $t=0$ but also all the streamlines in its neighborhood two vector equations should be integrated simultaneously:
\begin{eqnarray}
\frac{d\vec{x}}{dt} &=& \vec{v_t}, \mbox{  where  } \label{x-dot}\\
\vec{v_t} &=& - \frac{\partial h_t}{\partial y}\hat{x} + \frac{\partial h_t}{\partial x} \hat{y},  \mbox{ and  } \\
\frac{d\vec{\delta}}{dt} &=& \vec{\delta}\cdot\vec{\nabla}\vec{v_t}. \label{delta-dot}
\end{eqnarray}

Equation (\ref{x-dot}) is to be solved with the initial condition $\vec{x}(0) = x_0\hat{x} + y_0\hat{y}$, so solving that equation means solving two coupled equations, one for $dx/dt$ and one for $dy/dt$.  

Equation (\ref{delta-dot}) for $\vec{\delta}$ is obtained from the equation for neighboring magnetic field lines, which solves the exact equation $d(\vec{x}+\vec{\delta})/dt=\vec{v}_t(\vec{x}+\vec{\delta},t)$, by taking the limit as $\big|\vec{\delta}\big|\rightarrow0$.  Equation (\ref{delta-dot}) should be solved for two different initial conditions.  The first solve is for $\vec{\delta}_x = \delta_{xx}\hat{x} + \delta_{xy}\hat{y}$ with the initial condition  $\delta_{xx}=1$ and $\delta_{xy}=0$.  The second solve is for $\vec{\delta}_y = \delta_{yx}\hat{x} + \delta_{yy}\hat{y}$ with the initial condition  $\delta_{yx}=0$ and $\delta_{yy}=1$.  Since Equation (\ref{delta-dot}) for the evolution of the separation $\vec{\delta}$ is linear, the initial separation can be taken to be unity without loss of generality.   

The Jacobian matrix for the starting point $(x_0,y_0)$, which is defined by
\begin{eqnarray}
\frac{\partial\vec{x}}{\partial\vec{x}_0} &\equiv& \left(\begin{array}{cc}\frac{\partial x}{\partial x_0} & \frac{\partial x}{\partial y_0} \\\frac{\partial y}{\partial x_0} & \frac{\partial y}{\partial y_0}\end{array}\right) \mbox{   is then  } \\
&=&\left(\begin{array}{cc} \delta_{xx} & \delta_{xy}  \\\ \delta_{yx}  & \delta_{yy} \end{array}\right).
\end{eqnarray}
The determinant of the Jacobian matrix, $\delta_{xx}\delta_{yy}- \delta_{yx}\delta_{xy}$, called the Jacobian, would be unity if there were no numerical errors.  This follows from Liouville's theorem of  Hamiltonian mechanics.  

The Frobenius norm of a matrix is the square root of the sum of the squares of the matrix elements and is also equal to the square root of the sum of the squares of the singular values of a Singular Value Decomposition (SVD) of the matrix.  The Jacobian of a matrix is the product of its singular values, and a $2\times2$ matrix has two singular values $\Lambda_u$ and $\Lambda_s$; by definition $\Lambda_u\geq\Lambda_s$.

Consequently, the Frobenius norm of the Jacobian matrix, $\| \partial\vec{x}/\partial \vec{x}_0\|$, gives the large singular value, $\Lambda_u$, of a Singular Value Decomposition (SVD) of the matrix $\partial\vec{x}/\partial\vec{x}_0$:  
\begin{eqnarray}
\left\| \frac{\partial\vec{x}}{\partial \vec{x}_0} \right\| &\equiv& \sqrt{ \delta_{xx}^2+\delta_{xy}^2+\delta_{yx}^2 +\delta_{yy}^2}; \nonumber\\ 
 & =& \sqrt{\Lambda_u^2+1/\Lambda_u^2} \label{Frob-exp}
\end{eqnarray}
since $\Lambda_s=1/\Lambda_u$. 

When the flow is chaotic, neighboring streamlines separate exponentially, and $\Lambda_u$ becomes exponentially large, which means $\Lambda_u$ is essentially equal to the Frobenius norm of the Jacobian matrix, $\| \partial\vec{x}/\partial \vec{x}_0\|$.  A full SVD analysis gives additional information, the directions in both $x_0,y_0$ space and in $x,y$ space in which trajectories exponentiate apart and exponentiate together. 

The numerical accuracy of the calculations, which are based on Runge-Kutta integrations, can be checked not only by the deviation of the Jacobian from unity, but also by simultaneously integrating one additional equation, $dh_t/dt = \partial h_t/\partial t$, and finding the deviation of $h_t$ resulting from the integration from the actual $h_t$. 

The Frobenius norm of the Jacobian matrix is used here to define the magnitude of the exponentiation.  This norm  involves a sum of positive numbers and is less numerically demanding than calculating the SVD or the Jacobian, which is the difference between two numbers, each of order the Frobenius norm squared.  The largest Frobenius norm in this paper is approximately $10^8$.  The Jacobian can be the difference between two terms each of order $10^{16}$.   The maximum error in the Jacobian is 15\%.  A more representative number is the standard deviation of the Jacobian from unity, which is 1.1\%.


\subsection{Numerical results for the streamlines of $\vec{v}_t$ \label{sec:v-results} }

\begin{figure*}
\centerline{ \includegraphics[width=7in]{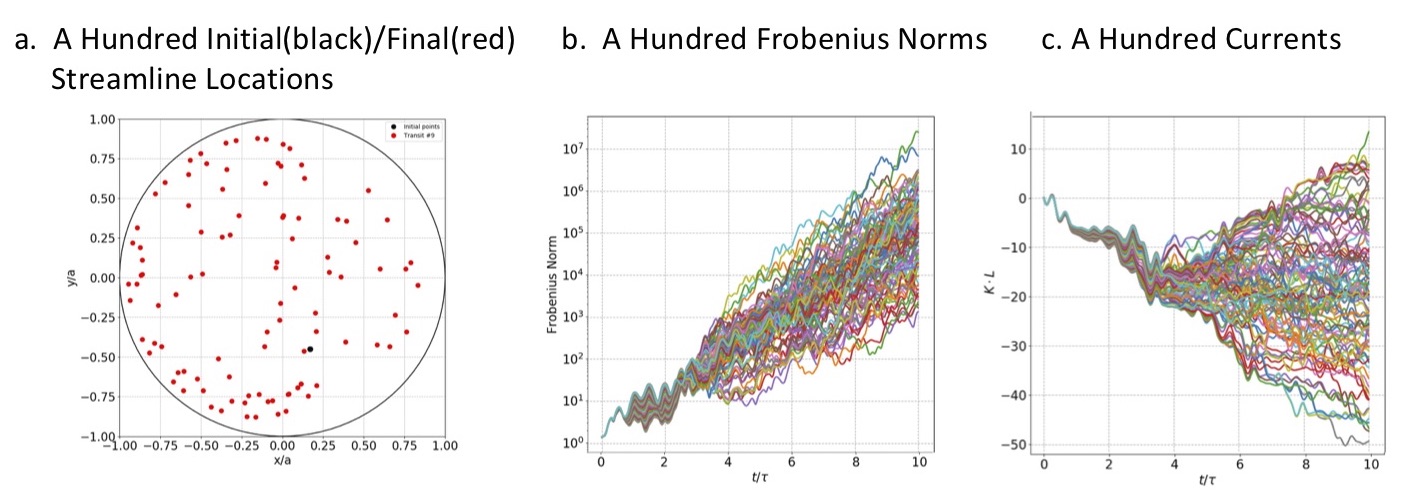}}
\caption{Streamline properties are illustrated for the stream function of Equation (\ref{h-tilde}) with $c_0=0$, $c_1=c_2=c_3=1/4$, $\omega_1 = 6\pi$, $\omega_2 = 4\pi$, $\omega_3 = 0$, and $\lambda^2=0$.  Figure \ref{fig:streamline}a is plot of a hundred streamlines started on the perimeter of the small black circle, which has a radius of $a/100$ and is centered at an arbitrary point, $x/a=0.17$ and $y/a=-0.45$.  The red dots are the locations of the hundred streamlines after nine transits.  The locations are widely scattered within the region $r<a$ in which the derivatives of the stream function are non-zero.  Figure \ref{fig:streamline}b shows the evolution of the Frobenius norm, Equation (\ref{Frob-exp}) for these hundred streamlines.  The Frobenius norm is a precise measure of the separation of neighboring trajectories.   As expected in a chaotic flow, the Frobenius norm tends to increase exponentially with time.  Figure \ref{fig:streamline}c shows the evolution of the force-free current $K\equiv \mu_0j_{||}/B$ times the length of the cylinder $L$ given by Equation (\ref{K-dot}) for each of the hundred field lines that initially intercepted the top surface on the perimeter of the black circle.  }  
\label{fig:streamline}
\end{figure*}

\begin{figure*}
\centerline{ \includegraphics[width=7in]{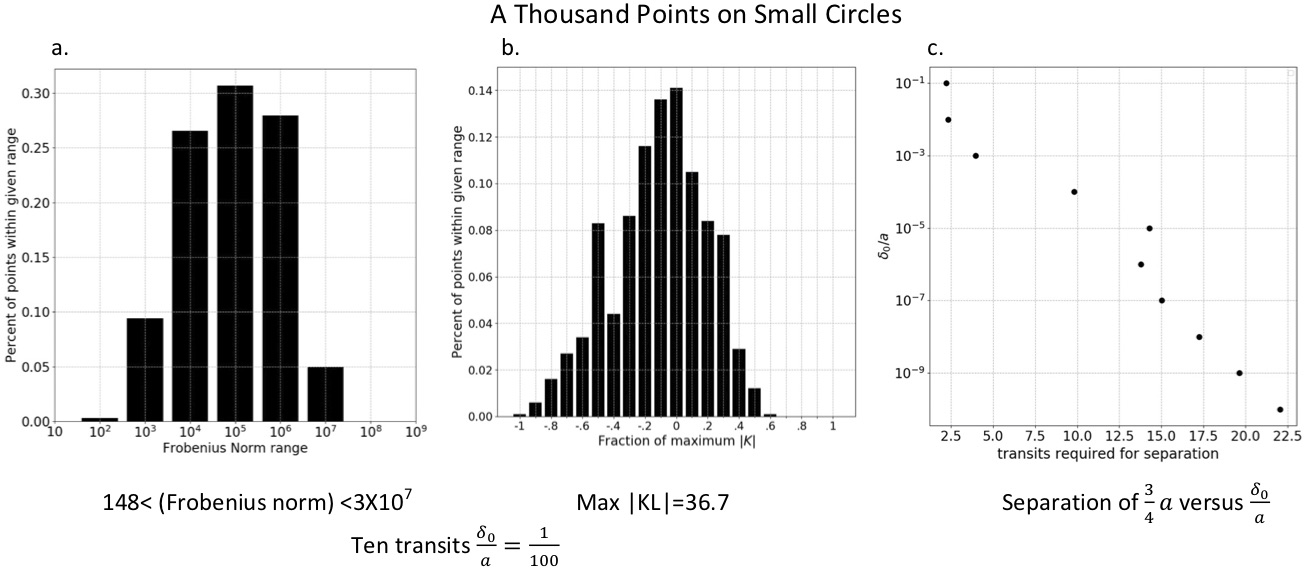}}
\caption{Data was gathered from a thousand starting points on the perimeter of a circle of radius $\delta_0$.  Figure \ref{fig:prob}a shows the frequency of occurrence of different values of Frobenius norm when $\delta_0=a/100$.  The same value as in Figure \ref{fig:streamline}.  Figure \ref{fig:prob}b shows the frequency of occurrence of different values of the current $K$, positive and negative, when $\delta_0=a/100$. Figure \ref{fig:prob}c shows the logarithmic scaling of the number of transits required for points started on the perimeter of a circle of radius $\delta_0$ to reach the scale $3a/4$.  Figures \ref{fig:prob}a and \ref{fig:prob}b used data from ten transits.   }  
\label{fig:prob}
\end{figure*}

\begin{figure*}
\centerline{ \includegraphics[width=7in]{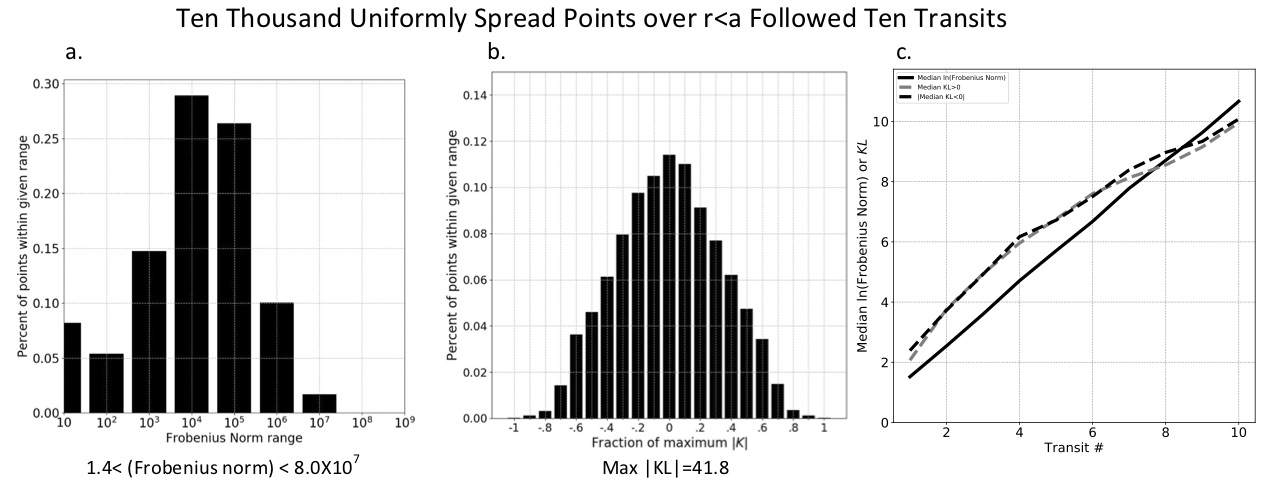}}
\caption{Ten thousand starting points were uniformly spread over $r<a$.  Figure \ref{fig:FullCircle}a is the frequency with which streamlines have various values of the Frobenius norm.  Figure \ref{fig:FullCircle}b is the distribution of the current, both positive and negative, relative to the maximum $\big| KL \big|=41.8$.  Figure \ref{fig:FullCircle}c relates the median value of the natural logarithm of the Frobenius norm (solid line), the median value the positive $KL$ magnitudes, and the median value of the negative $KL$ magnitudes. }  
\label{fig:FullCircle}
\end{figure*}

Figure \ref{fig:streamline}a illustrates the effect of a simple $\vec{v}_t$ that is chaotic, as almost all choices of $h_t$ that have non-trivial $x$, $y$, and $t$ dependencies are.  One hundred streamlines are started on the perimeter of the small black circle, but as the red dots illustrate, these hundred streamlines spread over most of the region $r<a$ after only five periods $\tau$ of the flow.   The perimeter of the small circle defines a tube in time with fixed area.  The cross section of this tube becomes convoluted in the extreme; all the red dots must lie on the perimeter of the tube.  Figure \ref{fig:streamline}b illustrates the evolution of the Frobenius norms of these hundred streamlines.  Figure \ref{fig:prob}a gives the frequency distribution of the Frobenius norms for a thousand streamlines started on the small black circle.  The smallest Frobenius norm for these thousand streamlines is 148 and the largest is $3\times10^7$.  The frequency distribution is peaked near their geometric mean $\sqrt{(148)\times(3\times10^7)}\approx6.7\times10^4$. 

A thousand streamlines were launched from a circle of radius $\delta_0$ and the number of transits, $t/\tau$, was recorded for the streamlines to become separated by a distance $3a/4$.  Figure \ref{fig:prob}c shows that the required number of transits is proportional to the logarithm of the chosen $\delta_0$.

The fraction of the total $r<a$ region that has chaotic streamlines is assessed by starting a thousand streamlines at uniformly spread points over the $r<a$ region and following them for ten transits.  The frequency with which various Frobenius norms arose is shown in Figure \ref{fig:FullCircle}a.   The smallest Frobenius norm is $\sqrt{2}$, which is the smallest value that is mathematically allowed.   The peak at small Frobenius norms implies that non-chaotic regions exist, which are separated by Lagrangian coherent structures from the chaotic regions \cite{Borgogno:2011}.   In non-chaotic regions, the separation between neighboring streamlines typically increases in proportion to time.  The frequency distribution of Frobenius norms is peaked near the geometric mean, $1.1\times10^4$, of the largest, $8\times10^7$, and the smallest Frobenius norm.    

 The fractional distribution of exponentiations in the separation of magnetic field lines was calculated in 2014 for a related problem by Huang et al \cite{Huang:2014} with far fewer e-folds; their distribution was also peaked.


\section{Evolution equations  \label{sec:evolution}   }

The magnetic field evolution in the model described in Section \ref{sec:model} becomes remarkably simple when the height of the cylinder is far greater than its radius, $L/a\rightarrow\infty$ and the time scale of ideal evolution is very long compared the the Alfv\'en transit time.   For simplicity the plasma pressure is assumed to be zero.  The derivation is essentially a simplified version of the derivation of Reduced MHD \cite{Kadomtsev,Strauss}.

This section consists of four subsections. The first two subsections,  Section \ref{sec:B-simp} and \ref{sec:force balance}, derive two differential equations that relate the Lagrangian derivatives of time and distance along the magnetic field lines $\ell$ of the distribution of parallel current $K\equiv \mu_0j_{||}/B$ and the vorticity $\Omega\equiv\hat{z}\cdot\vec{\nabla}\times\vec{u}_\bot$ of the magnetic field line flow.  The third subsection, Section \ref{sec:implications}, discusses the implications of these two equations and is the most important part of the paper.  The fourth subsection, Section \ref{sec:req-K}, gives a heuristic argument that the magnitude of the current distribution $K$ should typically scale as the strength of the exponential separation of the magnetic field lines, their Frobenius norm.  This relation is shown to hold accurately for the ensemble of many field lines but not for each line.  Indeed, the correlation between large values of $K$ and large values of the Frobenius norm is found to be remarkably weak.  


\subsection{Simplification of the magnetic evolution  \label{sec:B-simp} }  

When $L/a\rightarrow\infty$, the magnetic field consists of a constant field $B_0$ in the $\hat{z}$ direction plus an orthogonal field produced by the velocity in the top surface.   $\vec{\nabla}\cdot\vec{B}=0$  implies the magnetic field and the vector potential have the forms
\begin{eqnarray} 
\vec{B} &=& B_0\Big(\hat{z} + \hat{z}\times\vec{\nabla}H\Big); \label{B-eq} \\
\vec{A} &=& B_0\Big( \frac{\hat{z}\times \vec{x}}{2}- H\hat{z} \Big), \label{A}
\end{eqnarray} 
where $\vec{B}=\vec{\nabla}\times\vec{A}$.

Equation (\ref{Ideal ev}) for the ideal evolution of the magnetic field implies the vector potential evolves as 
\begin{eqnarray}
\frac{\partial \vec{A}}{\partial t}=\vec{u}_\bot\times\vec{B} - B_0\vec{\nabla}h,  \label{?A/?t}
\end{eqnarray}
where $B_0=\hat{z}\cdot\vec{B}$ is a constant and $h$ represents the freedom of gauge.  The constraint that the $\hat{z}$-directed field does not change is 
\begin{eqnarray}
&&\hat{z}\cdot\frac{\partial \vec{B}}{\partial t} = - \vec{\nabla}\cdot\Big(\hat{z}\times\frac{\partial \vec{A}}{\partial t}\Big) = 0. \label{B_0 const} 
\end{eqnarray}A vector identity implies $\hat{z}\times(\vec{u}_\bot\times\vec{B})=(\hat{z}\cdot\vec{B})\vec{u}_\bot - (\hat{z}\cdot\vec{u}_{\bot})\vec{B}$.  Consequently, the constraint on the constancy of $B_0=\hat{z}\cdot\vec{B}$ is that the velocity of the magnetic field lines have the form
\begin{eqnarray} 
\vec{u}_\bot &=& \hat{z}\times\vec{\nabla}h. \label{u-bot eq}
\end{eqnarray}

The curl of the magnetic field of Equation (\ref{B-eq}) and the curl of the magnetic field line velocity $\vec{u}_\bot$ of Equation (\ref{u-bot eq}) give the current and the vorticity along $\vec{B}$:
\begin{eqnarray}
&& \nabla_\bot^2 H = K,  \mbox{   where  } K\equiv \frac{\mu_0 j_{||}}{B}, \mbox{   and  } \label{K}\\ 
&& \nabla_\bot^2 h = \Omega,  \mbox{   where  } \Omega\equiv \hat{z} \cdot \vec{\nabla}\times\vec{u}_\bot. \label{Omega}
\end{eqnarray}


Equations (\ref{A}) and (\ref{?A/?t}) give two expressions for $\vec{B}\cdot(\partial\vec{A}/\partial t)_{\vec{x}}$, which can be equated to obtain
\begin{eqnarray}
\frac{\partial H}{\partial t} &=& \frac{\vec{B}\cdot \vec{\nabla}h}{B_0}\\
&=& \frac{\partial h}{\partial z} + (\hat{z}\times\vec{\nabla}_\bot H) \cdot \vec{\nabla}_\bot.
\end{eqnarray}
Since the magnetic field line velocity $\vec{u}_\bot = \hat{z}\times\vec{\nabla}h$,
\begin{equation}
\frac{\partial H}{\partial t} + \vec{u}_\bot\cdot\vec{\nabla}H =  \frac{\partial h}{\partial z}. \label{H ev}
\end{equation}
This equation can be written in Lagrangian coordinates in which ordinary Cartesian coordinates $\vec{x}=x\hat{x} + y\hat{y}+z\hat{z}$ are given as $\vec{x}(\vec{x}_L,t)$ with
\begin{eqnarray}
\left(\frac{\partial \vec{x}}{\partial t}\right)_L &=& \vec{u}_\bot,  \mbox{   so  }\\
\left(\frac{\partial H}{\partial t}\right)_L&=& \left(\frac{\partial H}{\partial t}\right)_{\vec{x}} + \frac{\partial H}{\partial \vec{x}}\cdot \left(\frac{\partial \vec{x}}{\partial t}\right)_L \\
&=&\frac{\partial H}{\partial t} + \vec{u}_\bot\cdot\vec{\nabla}H. \mbox{   Consequently  } \hspace{0.2in} \\
\left(\frac{\partial H}{\partial t}\right)_L&=& \frac{\partial h}{\partial z} \label{Lag H ev}
\end{eqnarray}
using Equation (\ref{H ev}).

Although the form is more complicated involving the metric tensor, Laplacians can be calculated in Lagrangian coordinates, and the relations $\nabla_\bot^2H=K$ Equation (\ref{K}), and $\nabla_\bot^2h=\Omega$, Equation (\ref{Omega}), remain valid.  Applying $\nabla_\bot^2$ to both sides of Equation (\ref{Lag H ev}), 
\begin{equation}
\left(\frac{\partial K}{\partial t}\right)_L = \left(\frac{\partial \Omega}{\partial \ell}\right)_L,   \label{?K/?t} 
\end{equation}
where the subscript $L$ on the partial derivatives implies the use of Lagrangian coordinates, which means $x_0$ and $y_0$ are held constant;
\begin{eqnarray}
&&\left(\frac{\partial K}{\partial t}\right)_L \equiv \frac{\partial K}{\partial t} + \vec{u}_\bot\cdot\vec{\nabla}K, \mbox{     and   }\\
&&\left(\frac{\partial \Omega}{\partial \ell}\right)_L \equiv \vec{B}\cdot\vec{\nabla}\Omega.
\end{eqnarray}
The differential distance along a magnetic field line, $d\ell$, is equivalent to $dz$ with $x_0$ and $y_0$ held constant.  

The implications of Equation (\ref{?K/?t}) will be found to be extremely profound.

\subsection{Constraint of force balance \label{sec:force balance}}

The Lagrangian time and $\ell$ derivatives of the parallel current distribution $K\equiv\mu_0j_{||}/B$ and the vorticity $\Omega$ obey not only Equation (\ref{?K/?t}) but also Equation (\ref{force}), which is implied by force balance.

For simplicity, the plasma is assumed to have a negligible pressure and a constant density $\rho$, so force balance is $\rho (\partial \vec{u}_\bot/\partial t + \vec{u}_\bot\cdot\vec{\nabla}\vec{u}_\bot)= \vec{f}_L$, where $\vec{f}_L \equiv\vec{j}\times\vec{B}$ is the Lorentz force.  The condition $\vec{\nabla}\cdot\vec{j}=0$ can be written as
\begin{eqnarray}
 \vec{B}\cdot\vec{\nabla}K &=& \vec{B}\cdot\vec{\nabla}\times\frac{\mu_0\vec{f}_L}{B^2},  \mbox{  and  } \label{parallell K deriv} \\
 \vec{\nabla}\times( \vec{u}_\bot\cdot\vec{\nabla}\vec{u}_\bot)&=&\vec{\nabla}\times(\vec{\Omega} \times \vec{u}_\bot)\\
 &=& \vec{u}_\bot\cdot\vec{\nabla}\vec{\Omega} - \vec{\Omega} \cdot  \vec{\nabla} \vec{u}_\bot,
 \end{eqnarray}
where $\vec{\Omega}=\Omega \hat{z}$.   The $\hat{z}$ component of the curl of the force balance equation gives
 \begin{eqnarray}
 \left(\frac{\partial\Omega}{\partial t}\right)_L = V_A^2  \left(\frac{\partial K}{\partial\ell}\right)_L, \mbox{  where   } V_A^2\equiv \frac{B_0^2}{\mu_0\rho} \label{force}
 \end{eqnarray}
 is the Alfv\'en speed.
 
 
 \subsection{Implications of the $K$ and $\Omega$ equations \label{sec:implications} }
 
Equations (\ref{?K/?t}) and (\ref{force}) together with the mixed partials theorem applied to either $\Omega$ or $K$ imply both $\Omega$ and $K$ obey the equation for shear Alfv\'en waves, $(\partial^2 K/\partial t^2)_L = V_A^2 (\partial^2 K/\partial \ell^2)_L $.  Any variation in $K$ along the magnetic field lines relaxes by Alfv\'en waves.  Reconnection or ideal kink-instabilities will generally drive Alfv\'en waves.  The inclusion of resistivity or viscosity  causes these waves to diffuse across the magnetic field lines and produces wave decay \cite{Boozer:j-||}.  In a completely ideal theory, the energy that goes into Alfv\'en waves will bounce back and forth forever, but they can be damped without directly affecting reconnection by adding viscosity or a drag-force to the force equation.
 
 Equations (\ref{?K/?t}) and (\ref{force}) imply that during any period in which the evolution is slow compared to the Alfv\'en transit time $L/V_A$ that
 \begin{eqnarray}
&& \left(\frac{\partial K}{\partial \ell}\right)_L=0, \mbox{  and  } \\
 && \left(\frac{\partial^2 \Omega}{\partial \ell^2}\right)_L=0, \mbox{  so  } \\
&&  \Omega = \Omega_t(x_0,y_0,t)\frac{\ell}{L}, \mbox{   and  } \\
&&  \left(\frac{\partial K}{\partial t}\right)_L = \frac{\Omega_t(x_0,y_0,t)}{L}, \mbox{   where   } \label{K-dot} \\
 && \Omega_t \equiv \hat{z}\cdot \vec{\nabla}\times\vec{v}_t. 
 \end{eqnarray}
 The flow of the top perfectly-conducting surface is specified, and $\Omega_t(x_0,y_0,t)$ is obtained from that specified flow alone.  For an example of a calculation of the evolution of $KL$, see Figure \ref{fig:streamline}c.

 Equations (\ref{K}) and (\ref{K-dot}) provide a Poisson equation, $\nabla_\bot^2H=K$, for $H$ and an expression for $K$, which can be solved for each value of $z$ and $t$.  The boundary condition is that the component of $\vec{\nabla}_\bot H$ that is tangential to the wall must vanish, otherwise the magnetic field would penetrate the perfectly conducting wall.
 
 For any physically reasonable flow, $\Omega_t(x_0,y_0,t)$ is bounded, $\big| \Omega_t \big| \leq\Omega_{max}$, which can be easily calculated analytically for any analytic $h_t(x,y,t)$.  An extremely important result is that the maximum current density along the magnetic field $j_{||}$ satisfies
 \begin{eqnarray}
 K_{max} \leq \frac{\Omega_{max}}{L} t. \label{K-max}
 \end{eqnarray}

 The fraction of the values of $K$ that have a particular value is illustrated for magnetic field lines started on a small circle of radius $a/100$  in Figure \ref{fig:prob}b and for lines started uniformly over the full region, $r<a$ in Figure \ref{fig:FullCircle}b.  For the small circle, the current $K$ is more likely to be negative than positive, which is also clearly illustrated in Figure \ref{fig:streamline}c, but nonetheless $K$'s of both signs are present.   Over the full region, frequency distribution of currents, Figure \ref{fig:FullCircle}b, is essentially symmetric between the negative and the positive values, and the most probable $K$ is essentially zero.  The absence of smoothness in the current distribution $K$, even in a small region, Figure \ref{fig:streamline}c, due to a smooth flow may be surprising to some.  An implication is that currents in the corona must be extremely complicated with a short correlation distance across the magnetic field lines.
 
 When the specified flow in the top surface is chaotic, the spatial derivatives of $K$ will tend to become exponentially large in some directions and exponentially small in others, but the current density itself is strictly bounded by a linear increase in time.  In other words, the current density within the plasma lies in ribbons with a decreasing thickness in one direction across $\vec{B}$, an increasing width in the other direction across $\vec{B}$, and a constant amplitude along $\vec{B}$.  This thinning with increasing width is illustrated by the top row of Figure \ref{fig:transit}.  
 
 The anisotropy of the spatial derivatives of $K$ follows from the exponentially large anisotropy of the spatial derivatives of $\Omega_t(x_0,y_0,t)$ in $x_0,y_0$ space.  For simple stream functions $h_t(x,y,t)$, such as those defined by Equation (\ref{h-tilde}), $\Omega_t(x,y,t)=\nabla^2h_t$ has a simple and smooth variation in $x,y$ coordinates.  But, the streamlines, $x(x_0,y_0,t)$ and $y(x_0,y_0,t)$,  of a two-dimensional, divergence-free chaotic flow separate exponentially in time in one direction, which implies they must exponentially converge in the other.  For a divergence-free flow, the two singular values of the Jacobian matrix $\partial\vec{x}/\partial\vec{x}_0$ must be inverses of each other.   The spatial derivatives of $\Omega_t(x_0,y_0,t)$ in the converging direction become exponentially large and those in the diverging direction become exponentially small.  
 
Employing Equation (\ref{K-dot}), $\Omega_t(x_0,y_0,t)$ determines the distribution of the parallel current density $K(x_0,y_0,z,t)$.  Consequently, the streamlines of $\vec{v}_t$ determine the properties of $K(x_0,y_0,z,t)$ throughout the plasma.  In the limit $V_A>>L/\tau$, the Lorentz force and, therefore, $j_\bot$ are negligible.  Figure \ref{fig:streamline}c illustrates how a hundred magnetic field lines that initially had nearby $x_0,y_0$ locations develop a large variation in $KL$.  As can be seen in Figure \ref{fig:transit}, there is only a weak correlation between regions where $\big| K \big|$ is large and where the Frobenius norm is large.  This is consistent with the results of Reid et al \cite{Reid:2020} that the quasi-squashing factor, which is determined by the Frobenius norm, Equation (\ref{Frob-exp}), has a little correlation with a large $\int \eta j_{||} d\ell = B_0 (\eta/\mu_0)KL$.  
 
Magnetic field lines that have distant intersection points with the bottom surface of the cylinder, $x_0,y_0$ and $x'_0,y'_0$ can interchange their intersections on the top surface if anywhere along their trajectories they are sufficiently close, $<\Delta_d$, to be indistinguishable.   This means they come closer than $c/\omega_{pe}$ or the distance through which they resistively diffuse, $\sqrt{(\eta/\mu_0)t}$.


\subsection{Required current density \label{sec:req-K} }
 
\begin{figure*}
\centerline{ \includegraphics[width=7in]{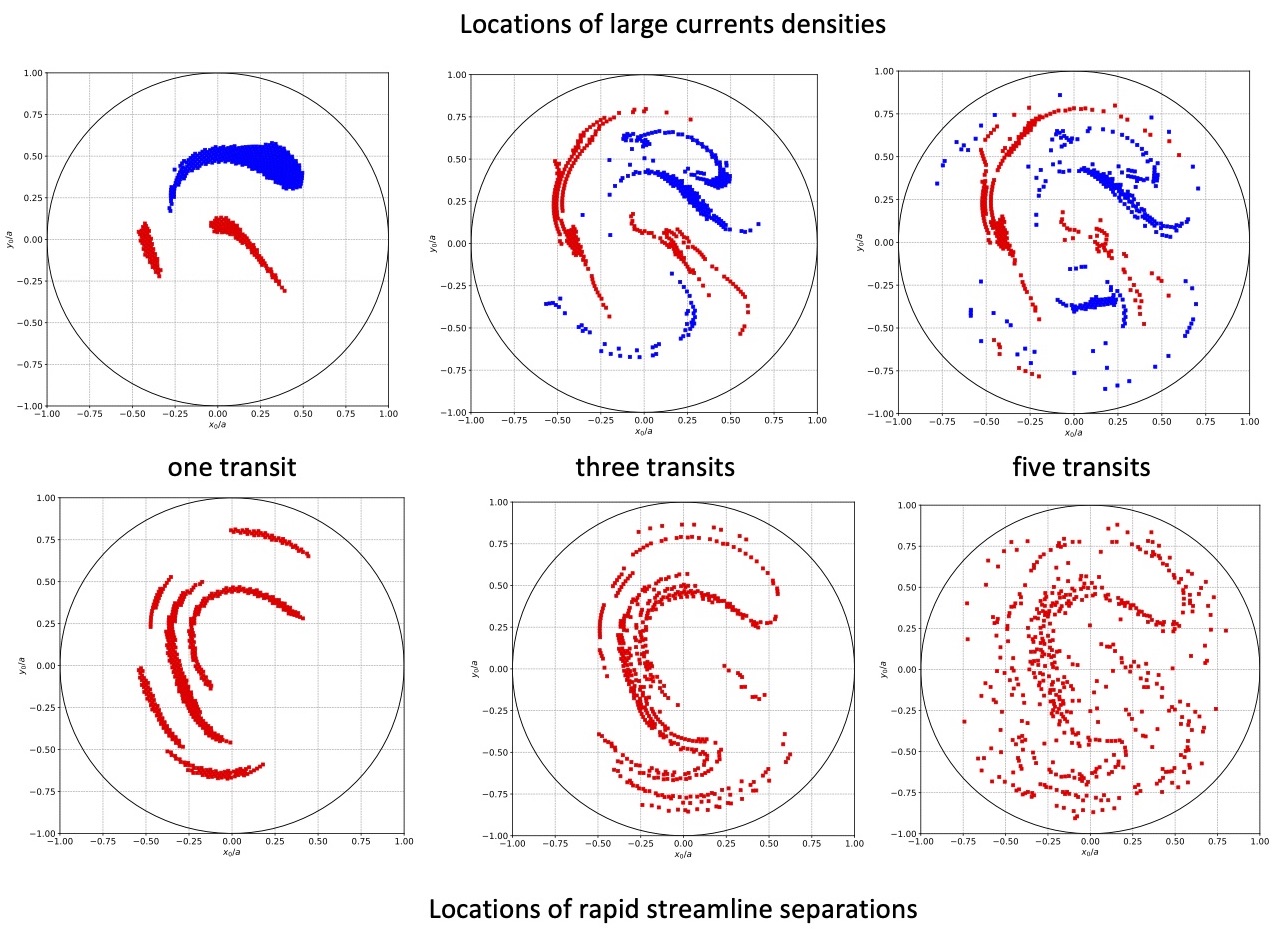}}
\caption{Ten thousand starting points were uniformly spread over $r<a$.  The top row is the locations of the five hundred points that had the largest current density $\big| K \big|$ after one, three, and five transits.  Red implies $K$ is negative and blue positive.    The bottom row is the locations of the five hundred points that had the largest Frobenius norm, which measures the rapidity of streamline separation.  The correlation between a large $K$ and a large Frobenius norm is weak.  The regions with a high current density $K$ tend to become long and thin, but small regions that have a correlated current density do not entirely disappear.  Figure \ref{fig:streamline}c also illustrates this.  }  
\label{fig:transit}
\end{figure*}

 The current density required for a large exponentiation is relatively small \cite{Boozer:B-line.sep}.  The minimum number of exponentiations is given by the properties of $h_t$, but more are possible.   The separation $\vec{\Delta}=\Delta_x\hat{x} +\Delta_y\hat{y}$ between two neighboring magnetic field lines obeys $d\vec{\Delta}/dz = \vec{\Delta}\cdot \vec{\nabla}\hat{b}$, where $\hat{b} = \hat{z}+\hat{z}\times\vec{\nabla}H$.  That is,
 \begin{eqnarray}
 \frac{d\Delta_x}{dz} &=& - \frac{\partial^2 H}{\partial x\partial y} \Delta_x - \frac{\partial^2 H}{\partial y^2} \Delta_y \label{Delta_x} \\
  \frac{d\Delta_y}{dz} &=&  \frac{\partial^2 H}{\partial x^2} \Delta_x  + \frac{\partial^2 H}{\partial x\partial y} \Delta_y. \label{Delta_y}
  \end{eqnarray}
An exact answer for the separation requires a solution of the equation  $\nabla_\bot^2H = K$.  But, the typical magnitude of the second derivatives of $H$, which appear in Equations (\ref{Delta_x}) and (\ref{Delta_y})  is $K$, which suggests that the number of e-folds is typically of order $KL$.  Figure \ref{fig:streamline} illustrates this scaling, and Figure \ref{fig:FullCircle}c shows the accuracy with which the scaling holds.  Although the scaling holds for the ensemble averages, it does not hold magnetic field line by field line, which is another way of saying that the correlation between the magnitude of the Frobenius norm and the current density is weak, Figure \ref{fig:transit}.  A current along a magnetic field line affects not only the Hamiltonian and its derivatives on that line but elsewhere as well.

   \section{Kink stability \label{sec:kink} }
   
   The current flowing along the magnetic field lines causes the lines to twist through an angle $\Theta = KL/2$ from one end of the cylinder to the other.  When the twist has a smooth variation with radius, Hood and Priest \cite{Hood:kink1979} found the magnetic field becomes unstable to an ideal kink when $\Theta$ is greater than a critical value, which in their calculations lay in the range $2\pi$ to $6\pi$.   Studies of the onset of reconnection in the model of Figure \ref{fig:cylinder} are much simpler when ideal kink instabilities are not an issue. 
   
The largest $KL$ values in Figure \ref{fig:streamline}c correspond to $\Theta \approx 8\pi$, but the current $K$ has an extremely complicated spatial distribution, not only in magnitude but also in sign; spatial averages are  far smaller than the maximum value.   As will be discussed, the anisotropy of the derivatives of $K$ across the magnetic field lines and the smallness of the spatial averages of $K$ makes the system highly stable to kinks.

It is not required that $K$, or equivalently the twist $\Theta$, have small spatial averages when the flow is chaotic.  The spatially averaged twist $\Theta$ can be made arbitrarily large by choosing $c_0$ to be large and $\omega_0$ to be either zero or small in Equation (\ref{h-tilde}) for $\tilde{h}$.  The choice $c_0=0$ was made to show that a large average field line twist is not needed to obtain chaos.  

Even when $c_0=0$, the spatial average of $K$ over small regions can be non-zero.  This is illustrated by Figure \ref{fig:streamline}c and by the first row of Figure \ref{fig:transit}.

When the stream function is chosen so the flow is chaotic but with a large spatially-averaged $K$, the resulting magnetic field will generally evolve not only into a kinked but also into an eruptive state.  As shown in Section \ref{sec:helicity}, the evolution properties of magnetic helicity imply the spatial and temporal average of $h_t$ must be zero for a non-eruptive steady-state solution for the magnetic field when $R_m>>1$---no matter how spatially concentrated the current may become.  Consequently, non-eruptive chaotic models tend to have spatially complicated distributions of $\Theta$ in which $\Theta$ has both signs and a near-zero spatial average as in the $\vec{v}_t$ example used to construct Figure \ref{fig:streamline}.  

As discussed in Section V.B.1 of \cite{Boozer:RMP}, the stability of force-free equilibria can be determined using the perturbed equilibrium equation $\vec{\nabla}\times\delta\vec{B} =(\mu_0\delta j_{||}/B)\vec{B}$, where $\delta\vec{B}=\vec{\nabla}\times(\delta A_{||}\hat{z})$.  The perturbed parallel current is determined by the constancy of $ K\equiv\mu_0 j_{||}/B$ along magnetic field lines, which in linear order in the perturbation implies $\vec{B}\cdot\vec{\nabla} \delta K + \delta\vec{B}\cdot \vec{\nabla}K=0$.  Stability is determined by whether it takes positive or negative energy to drive a perturbation that obeys the equations
\begin{eqnarray}
\nabla_\bot^2 \delta A_{||} &=& - \delta K B_0;  \label{delta-A}   \\
B_0\left(\frac{\partial \delta K}{\partial \ell}\right)_{x_0y_0} &=&  \hat{z}\cdot\left(\vec{\nabla}_\bot A_{||} \times \vec{\nabla}_\bot K\right). \hspace{0.2in} \label{delta-K}
\end{eqnarray}
The system is at marginal stability when $\delta K$ is just strong enough to produce a solution $\delta A_{||}$ that fits within the perfectly conducting cylindrical walls.  The implication is that when Equation (\ref{delta-A}) is multiplied by $\delta A_{||}$, then at marginal stability
\begin{eqnarray}
\int \left\{ \left(\vec{\nabla}_\bot \delta A_{||}\right)^2 - \delta K \delta A_{||} B_0\right\}d^3x =0. \label{stab-int}
\end{eqnarray}
When $\delta K$ has only a rapid spatial variation, as it does when $K$ does, then $\delta A_{||} $ must also have a rapid variation to avoid a self-cancelation of the destabilizing term $\delta K \delta A_{||}$ in Equation (\ref{stab-int}).  Equation (\ref{delta-K}) for $\delta K$ involves two spatial derivatives across $\vec{B}$ and one might think they could be large and balance the stabilizing effect of the two spatial derivatives in $\left(\vec{\nabla}_\bot \delta A_{||}\right)^2 $, but that is not the case.  The two spatial derivatives  across $\vec{B}$ in the equation for $\delta K$ are orthogonal and the spatial derivatives of $K$ in the two directions across the magnetic field lines  tend to be of exponentially different magnitudes, so the large term in  $\vec{\nabla}_\bot K$ forces a large spatial derivative in $\delta A_{||}$, which quadratically enhances $\left(\vec{\nabla}_\bot \delta A_{||}\right)^2$  but only linearly enhances $\delta K$.

   
   \section{Magnetic flux \label{sec:flux}}
   
   The change in the magnetic flux associated with a particular magnetic field line $\psi(x_0,y_0,t)$ is the integral from one perfectly conducting surface to the other, $\partial \psi/\partial t=-\int \vec{E}\cdot d\vec{\ell}$.  When the cylindrical conductor is stationary and the plasma is resistive, the flux decays as $\partial \psi/\partial t=-\int \eta \vec{j} \cdot d\vec{\ell}$.  
   
 As was shown in the derivation of Equation (\ref{?A/?t}) for $\partial\vec{A}/dt$, the effective inductive electric field along the magnetic field is $-\vec{B}\cdot(\partial A/\partial t) = \vec{B}\cdot(B_0\vec{\nabla}h)$, which gives a change in the flux, $\partial \psi/\partial t=-\int \vec{E}\cdot d\vec{\ell}$, or
 \begin{equation}
 \frac{\partial\psi}{\partial t} = - B_0 h_t(x_0,y_0,t) \label{flux-creation}
 \end{equation}
 since the at the top of the cylinder $h=h_t$.
 
 The appearance of $B_0h_t$ in the electric-field integral can be understood using the expression $\vec{E}= - \vec{v}_t\times\vec{B}_0$ for the electric field in the flowing conductor when observed from a stationary frame of reference.  The velocity is $\vec{v}_t=\hat{z}\times\vec{\nabla}h_t$ and $\vec{B}_0=B_0\hat{z}$,  so $\vec{E}=-B_0\vec{\nabla}_\bot h_t$.  The electromotive force from the intersection point of the field line to a stationary point, where $h_t=0$, is $B_0h_t$.

 The rate at which plasma resistivity destroys magnetic flux is $\mathcal{E}_\eta=\int\eta j_{||}d\ell$.   Since $K=\mu_0j_{||}/B$ is constant along a magnetic field line, $\mathcal{E}_\eta=  B_0(\eta/\mu_0) K(x_0,y_0,t)L$.  Equation (\ref{K-dot}) implies $\partial \mathcal{E}_\eta/\partial t=
-B_0(\eta/\mu_0)\Omega_t(x_0,y_0,t)$.  Since  $\Omega_t=\nabla^2 h_t$, the ratio of flux creation to flux destruction is 
 \begin{equation}
 \Big| \frac{2\partial h_t/\partial t}{(\eta/\mu_0)\nabla^2 h_t} \Big| \approx R_m \sim 10^{12} \label{res.flux}
 \end{equation} 
 in the solar corona. 
 
 
 \section{Magnetic helicity \label{sec:helicity}}
 
As will be shown, an argument based on magnetic helicity implies that a long-term relevant solution to the problem outlined in Figure \ref{fig:cylinder} requires the long-term spatial and temporal average of $h_t$ to be zero.  When the average of $h_t$  is zero over a chaotic region, the interchange of penetration points implies the poloidal magnetic flux associated with a field line $x_0,y_0$ fluctuates but has no systematic increase.    

Chaotic streamlines can cause two field lines that penetrate the bottom of the cylinder at two distinct points $x_0,y_0$ and $x'_0,y'_0$ to interchange their penetration points through the top plane due to exponentially small non-ideal effects.  
 
Equation (\ref{helicity-resistivity}) for the evolution of the magnetic helicity limits the degree to which a magnetic field driven as in Figure \ref{fig:cylinder} can be simplified by magnetic field lines exchanging connections even if the current density were to obtain arbitrarily high local values by being concentrated in thin sheets.  As has been shown, the maximum current density increases only linearly in time, Equation (\ref{K-max}), and does not reach the enhancement by a factor of order $1/R_m$, which would be required for the loop voltage to balance the poloidal flux creation, before reconnection has already occurred.  But, even if it did the rate of helicity dissipation would not be significantly enhanced, Equation (\ref{helicity-resistivity}).  Magnetic turbulence can reduce the magnetic energy, but not the helicity \cite{Taylor:1974,Berger:1984} as $R_m\rightarrow\infty$.

Equation (\ref{helicity-dot}) for the rate of helicity increase implies that unless the stream function integrated over each chaotic region, $\int h_t da_t$,  has a zero time average, the magnetic helicity can increase without limit.  In the model of this paper, the perfectly conducting cylindrical boundary conditions will keep the system confined no matter how strong or contorted the magnetic field may become. But, in a natural system, such as the solar corona, a drive $h_t$ that does not have a zero long-term average will presumably cause the eruption of a magnetic flux tube.

The derivation of the helicity evolution equation starts with the definition of the magnetic helicity enclosed by the cylinder,
\begin{eqnarray}
\mathcal{K}&\equiv& \int \vec{B}\cdot\vec{A} d^3x.
\end{eqnarray}
Equations (\ref{B-eq}) and (\ref{A}) together with $\vec{\nabla}\cdot\vec{x}_\bot=2$ imply $\vec{B}\cdot\vec{A} = B_0^2 (-2H + \vec{\nabla}_\bot\cdot(H\vec{x}_\bot)$.  The helicity is then
 \begin{equation}
 \mathcal{K} =-2B_0^2 \int H d^3x. \label{helicity-H}
 \end{equation}
 The time derivative of the helicity is calculated using
\begin{eqnarray}
\frac{\partial\vec{B}\cdot\vec{A}}{\partial t} &=& \vec{B}\cdot\frac{\partial\vec{A}}{\partial t}+\vec{A}\cdot\vec{\nabla}\times\frac{\partial\vec{A}}{\partial t}\nonumber\\
&=& 2 \vec{B}\cdot\frac{\partial\vec{A}}{\partial t} - \vec{\nabla}\cdot \left(\vec{A}\times\frac{\partial\vec{A}}{\partial t}\right) \mbox{  and  }\\
\vec{A}\times\frac{\partial\vec{A}}{\partial t}&=& - B_0^2 \frac{\vec{x}_\bot}{2}\frac{\partial H}{\partial t},  \mbox{   so   }\\
\frac{\partial\vec{B}\cdot\vec{A}}{\partial t} &=&2 \vec{B}\cdot\frac{\partial\vec{A}}{\partial t} +\vec{\nabla}\cdot\left(B_0^2 \frac{\vec{x}_\bot}{2}\frac{\partial H}{\partial t}\right) \label{?AB/?t} 
\end{eqnarray}
The side of the cylinder is a rigid perfect conductor, so $\partial H/\partial t=0$ within its sides.  Consequently,
\begin{eqnarray}
\frac{d \mathcal{K}}{dt} &=& 2 \int \vec{B}\cdot\frac{\partial\vec{A}}{\partial t} d^3x \\
&=& -2B_0\int \vec{\nabla}\cdot (h\vec{B}) d^3x \\
&=& - 2 B_0^2 \int h_t da_t, \label{helicity-dot}
\end{eqnarray}
using Equation (\ref{?A/?t}) with $da_t = rdrd\theta$ the area integral over the top of the cylinder.   Since the magnetic field lines are tied to the flow of the perfectly conducting top, $h=h_t$ within the top surface.  

The effect of resistivity on the helicity is obtained by letting $\partial\vec{A}/\partial t = \vec{u}_\bot \times \vec{B}- B_0\vec{\nabla}h - \eta\vec{j}$, then Equation (\ref{helicity-dot}) implies
\begin{eqnarray}
\frac{d \mathcal{K}}{dt} &=& - B_0^2 \int \left(h_t + \frac{\int \eta j_{||} d\ell}{B_0}\right)da_t\\
&=& - B_0^2 \int \left(h_t + \int \frac{\eta}{\mu_0}K d\ell\right)da_t. \label{helicity-resistivity}
\end{eqnarray}
The effect of resistivity on the helicity evolution is given by the volume-averaged $K$, and is therefore unaffected by $K$ being concentrated.  When the evolution is slow compared to the Alfv\'en transit time, $K$ is independent of $\ell$, and the ratio of helicity input to its resistive destruction is
\begin{eqnarray}
\left| \frac{h_t}{ \int \frac{\eta}{\mu_0}K d\ell} \right| &\sim& \frac{av_t}{ \frac{\eta}{\mu_0} L \frac{v_t t}{a^2L}} \nonumber\\
&\sim& \left(\frac{\mu_0 a^2}{\eta}\right) \left(\frac{v_t}{a}\right) \frac{a/v_t}{t} \\
&\sim& R_m \frac{a/v_t}{t},
\end{eqnarray}
 while $R_m\sim 10^{12}$ in the corona.   To extreme accuracy, resistivity has no effect on the rate of helicity increase in the corona.

 


\section{Power input \label{sec:power} }

The condition $\vec{\nabla}\cdot\vec{j}=0$ implies that the Lorentz force $\vec{f}_L\equiv\vec{j}\times\vec{B}$ obeys Equation (\ref{parallell K deriv}).  Using
\begin{eqnarray}
\vec{B}\cdot\vec{\nabla}K&=& B_0 \frac{\partial K}{\partial \ell} = -K\delta(\ell-L)\\
&=& B_0 \frac{\mu_0}{B_0^2} \hat{z}\cdot \vec{\nabla}\times \vec{f}_L,  \\
 \hat{z}\cdot \vec{\nabla}\times \vec{f}_L&=&-\frac{B_0^2}{\mu_0} K \delta(\ell-L).
\end{eqnarray}

The power required to maintain the flow in the top of the cylinder is
\begin{eqnarray}
\mathcal{P} &=&- \int \vec{v}_t\cdot \vec{f}_L d^3x \\
&=&-\int \hat{z} \cdot (\vec{\nabla}h_t \times \vec{f}_L) d^3x \nonumber \\
&=&- \int \hat{z} \cdot \Big(\vec{\nabla}\times(h_t \vec{f}_L) - h_t \vec{\nabla}\times\vec{f}_L\Big)d^3x  \nonumber\\
&=&- \int \hat{z} \cdot \Big(-\vec{\nabla}(h_t \hat{z} \times\vec{f}_L) \nonumber\\ && \hspace{0.2in} +  \frac{B_0^2}{\mu_0} h_t K\delta(\ell-L)\Big) d^3x  \nonumber\\
&=&-\frac{B_0^2}{\mu_0} \int h_t(x_0,y_0,t) K(x_0,y_0,t) dx_0dy_0. \label{Power} \hspace{0.1in}
\end{eqnarray}
The integrand becomes extremely spatially complicated as time evolves but is not very large.


 \section{Discussion \label{sec:discussion} }

 The prevalence of magnetic reconnection in situations in which effects that cause a departure from an ideal evolution are arbitrarily small suggests the cause of reconnection must be within the ideal evolution equation itself.  Indeed it is \cite{Boozer:ideal-ev}.  
 
 In an ideal evolution, the magnetic field lines move with the velocity $\vec{u}_\bot$ of Equation (\ref{Ideal ev}).  As Schindler, Hesse, and Birn \cite{Schindler:1988} stated in 1988, resistivity $\eta$ can only compete with the ideal evolution in a region $\Delta_d$ that is sufficiently narrow across the magnetic field  that the local magnetic Reynolds number, $R_\ell = (\mu_0 u_\bot/\eta) \Delta_d\sim1$.  The actual scale of the reconnecting region across the magnetic field, $a$, gives the usual magnetic Reynolds number $R_m = (\mu_0 u_\bot/\eta)a$, which is many orders of magnitude larger than unity---in both natural and laboratory plasmas.  In the solar corona $R_m\sim10^{12}$.  There are two possibilities for addressing the $a/\Delta_d\sim R_m$ problem identified by Schindler et al.  
 
 The first possibility, which is the dominant reconnection paradigm and the only one considered by Schindler et al, is that the ideal magnetic evolution creates and maintains layers of intense current density $j\sim B_{rec}/\mu_0 \Delta_d$, where $B_{rec}$ is the reconnecting magnetic field.   One problem with this possibility is that in an ideal evolution the current density tends to increase only linearly in time.  A linear increase in the current density by a factor $R_m$ takes far too long to explain many natural phenomena.  The growth in current density is found to be linear in time, not only in this paper, Equation (\ref{K-max}), but also in ideal flows that are known to create a singular current density as time goes to infinity---flows that have a resonant interaction with a rational magnetic surface in a torus \cite{Hahm-Kulsrud,Boozer-Pomphrey}.  Although more complicated than a linear increase, a slow increase in current density along the magnetic field lines is also found for lines that pass near a magnetic null \cite{Elder-Boozer}.
 
 What has been presented in this paper is an example of the second possible explanation for fast magnetic reconnection, but one that has aroused little interest.  This explanation is based on the characteristic increase in the ratio of the maximum to the  minimum separation, $\Delta_{max}/\Delta_{min}$, between two magnetic lines that is produced by an ideal flow $\vec{u}_\bot$.  Magnetic field lines are defined at a fixed time, as is $\Delta_{max}/\Delta_{min}$, but when the two lines are adjacent at some location along their trajectories, $\Delta_{min}\rightarrow0$, the $\Delta_{max}/\Delta_{min}$ ratio characteristically increases exponentially when the field-line trajectories are calculated at different points in time.  Characteristically, the rate of exponentiation is comparable to the evolution time $\tau_{ev}\equiv a/u_\bot$.    Reconnection must have occurred by the time at which $\Delta_{max}/\Delta_{min}\sim R_m$ with $\Delta_{min}=\Delta_d$, the spatial scale over which field line distinguishability is lost, and $\Delta_{max}\sim a$, the system scale.

 All that is required to produce reconnection by the Parker and Krook definition \cite{Parker-Krook:1956}, the ``\emph{severing and reconnection of lines of force}," is that magnetic field lines become indistinguishable on some spatial scale $\Delta_d$ and that the exponentiation of field-line separation magnify the indistinguishability scale to the scale over which reconnection occurs.  Indeed, one can model magnetic reconnection using an ideal code \cite{Pariat-Antiochos} because finite numerical resolution will provide an effective distinguishability scale $\Delta_d$.  In many examples of reconnection, the distinguishability scale $\Delta_d$ is far smaller than can be directly assessed numerically.  It is important to carry out reconnection studies minimizing effects that break the ideal magnetic evolution.
 
Unlike a localized current density, $\Delta_{max}/\Delta_{min}$ becomes large over extended regions.  Nonetheless, the exponentiation, which is quantified by the Frobenius norm, Section \ref{sec:quantification of chaos}, differs from one pair of magnetic field lines to another---even over small regions, Figure \ref{fig:streamline}b.  Magnetic field lines that have the largest exponentiations reconnect first, which can break force-balance and cause Alfv\'enic relaxations when the reconnected region becomes sufficiently large,  which is the scale $\Delta_r$ at which reconnection becomes important to the system dynamics.  The scale $\Delta_r$ is important but not determined in this paper.  That would require more complete simulations related to those of Reid et al \cite{Reid:2018}, and $\Delta_r$ is far less well defined than the distinguishability scale $\Delta_d$.  Nonetheless, $\Delta_r$ is an important concept because the typical size of reconnected regions scale as $\Delta_d$ times a factor that depends exponentially on time until the scale of reconnection reaches  $\Delta_r$, then reconnection spreads at an Alfv\'enic rate, which presumably accounts for Parker's observation \cite{Parker:1973}  that the speed of reconnection is $\approx0.1V_A$.  Even though an Alfv\'enic relaxation is consistent with an ideal evolution, it generally causes an  increase in $\Delta_{max}/\Delta_{min}$ on the Alfv\'enic time scale for pairs of magnetic field lines in additional regions of space.
 
 Plasma turbulence can be considered to be third possibility for fast reconnection, but turbulence-enhanced reconnection is in effect given by one of the other two possibilities.  Plasma turbulence can mean either micro-turblence or macro-turbulence: (1)  Micro-turbulence affects the particle distributions of the Fokker-Planck equation and can cause an enhanced turbulent resistivity $\eta_{turb}$ for the flow of current along $\vec{B}$, which can be much larger than the standard Spitzer value for the parallel resistivity, $\eta$.  This effect fits into the paradigm of Schindler, Hesse, and Birn \cite{Schindler:1988}; just $\eta_{turb} j_{||}$ must compete with the evolution rather than $\eta j_{||}$.  (2) Macro-turbulence means the mass flow velocity of the plasma $\vec{v}$ develops small spatial scales with no direct effect on the particle distribution function.  Macro-turbulence is analogous to the turbulent flow of water through a pipe.  
 
 In 1953 Dungey \cite{Dungey:1953} stated that a change in the linkages of magnetic field lines due to resistivity ``\emph{is usually very slow in astrophysical systems, but may be increased by turbulent motion in the gas.}"  Macro-turbulent enhanced reconnection has the same relation to enhanced reconnection due to the exponentiation of field line separations as the pre-Aref theories of fluid mixing to that of Aref \cite{Aref;1984}.  When macro-turbulence is three dimensional, it does cause the magnetic field to become chaotic.  Nevertheless when the characteristic spatial scale of the turbulent eddies $\ell_{turb}$ is small compared to the scale $a$ over which strong advection takes place, the advection from the eddies behaves as a diffusive process \cite{Boozer:rec-phys}.  The diffusive advection of the turbulent eddies is much slower than the advective transport due to flows that have a scale comparable to $a$ unless the maximum flow speed is far larger than the flow speed averaged over the scale $\ell_{turb}$.  A  deterministic  turbulent ideal flow $\vec{u}_\bot$ does not directly produce magnetic reconnection any more than a laminar flow does; both cause an exponentially enhanced sensitivity to non-ideal effects.
 

 The literature on macro-turbulence enhanced reconnection is extensive.  However, the focus of these studies was not on on deterministic magnetic fields and flows.  The effect on reconnection of indeterminacy in the magnetic field, which means intrinsic magnetic stochasticity,  was described in \cite{Lazarian:1999,Eyink:2011,Eyink:2015,Kowal:2020}, and reviewed in \cite{Lazarian:2020rev}.  The effect of turbulence and intermittency of plasma flows on two-dimensional magnetic fields was described in \cite{Matthaeus1986,Matthaeus:2015,Matthaeus:2020}.

The simple model developed here illustrates how the laminar flow of an ideal magnetic evolution can force magnetic reconnection on a time scale that is the ideal evolution time, $\tau_{ev}\equiv a/u_\bot$ times the logarithm of the ratio of the intrinsic magnitude of the ideal to the connection breaking terms. 

A spatially bounded flow of a footpoint of a magnetic field line that has a smooth and slow variation in space and time can be chaotic and impart properties to the magnetic field lines in an ideal evolution that many may find to be surprising.     

Two magnetic field lines that are separated by the distance $\Delta_{min}$ at the stationary foot point must have a maximum separation $\Delta_{max}$ at least as great as the separation of the lines at the moving footpoint.  The separation of neighboring streamlines determines the minimum $\Delta_{max}/\Delta_{min}$ ratio that the two magnetic field lines must have obtained.  Figure \ref{fig:FullCircle}c shows that no matter how small $\Delta_{min}$ may be that $\Delta_{max}$ will reach a value comparable to $a$, the distance scale of the system perpendicular to the magnetic field, within a time that depends only logarithmically on $\Delta_{min}$.  Identifying $\Delta_{min}$ with the spatial scale $\Delta_d$ on which magnetic field lines become indistinguishable, the time required for large-scale reconnection to occur is the evolution time defined by the footpoint flow speed times a logarithmic factor depending on $a/\Delta_d$.  This argument assumes the reconnection scale $\Delta_r$ is comparable to $a$, although it could be smaller, which would imply large scale reconnection would occur even sooner.

Many more properties of an ideally evolving magnetic field can be determined from the footpoint motion when its evolution time is long compared to the time for an Alfv\'en wave to propagate from one footpoint interception to the other and the magnetic field remains kink stable.  The example of the footpoint flow that was studied was chosen to be consistent with the maintenance of kink stability.  Two properties that can be determined are the distribution of the current density $K\equiv \mu_0j_{||}/B$, which is constant along the magnetic field lines under these conditions, and the power that is required to drive the footpoint flow.  The figures show the quantity $KL$ rather than $K$ because $KL$ is the quantity that is determined by the theory; not $K$ and $L$ separately.  $KL/2$ is the dimensionless twist angle that a magnetic field line makes in going between the two footpoints of each field line, which are separated by the distance $L$.  

Despite the smoothness and simplicity of the footpoint flow that was studied, the current density distribution $K$ varies wildly among field lines that have their stationary foopoints in a tiny region, Figure \ref{fig:streamline}c.  The currents in the corona must be extremely complicated with a short correlation distance across the magnetic field lines.  Observations of spatial complexity in the current density are clearly not a proof of turbulence.   Nearby magnetic field lines can have currents flowing in opposite directions.  In addition  the magnitude of $K$ increases only linearly in time, Figure \ref{fig:streamline}c, and forms ribbons along the magnetic field  that tend to widen exponentially and thin as one over the exponential, Figure \ref{fig:transit}.   As discussed in Appendix E of \cite{Boozer:part.acc}, the currents produced by the more localized photospheric motions could produce runaway electrons and explain the hot corona.

To be consistent with Ampere's law, $K$ must have a typical magnitude that is proportional to the logarithm of the exponentiation factor  \cite{Boozer:B-line.sep}.  A heuristic argument is given in Section \ref{sec:req-K}.  The exponentiation factor is the Frobenius norm of a Jacobian matrix, Section \ref{sec:quantification of chaos}.  This approximate proportionality is illustrated in Figure \ref{fig:FullCircle}c and requires the current density be enhanced by factor of $\ln(R_m)$ above its characteristic value before reconnection due to exponentially enhanced resistive diffusion can compete with evolution.  An enhancement of the current density by $\ln(R_m) \sim \ln(10^{12}) \approx 28$ is modest compared to an enhancement by $R_m\sim10^{12}$.  Despite this relationship between $K$ and the Frobenius norm, the regions in which they are particularly large are not closely correlated.  Reid et al \cite{Reid:2020} also found a weak correlation between the current density $j_{||}$ and ratio at the two footpoint locations of the distance between neighboring field lines, which is the definition of the quasi-squashing factor \cite{Priest:1995,Titov: 2002}.

The power required to drive the footpoint motion, Section \ref{sec:power}, is a spatial average of the footpoint motion times the current distribution $K$, so it increases approximately linearly in time and never becomes extremely large.

Section \ref{sec:helicity} demonstrated that unless the stream function $h_t(x,y,t)$, which defines the footpoint flow, has a zero spatial and temporal average over a footpoint region, the magnetic helicity increases without limit in the volume occupied by the magnetic field lines that strike that region.  When $R_m>>1$, magnetic helicity cannot be destroyed faster that it is created in such regions---even by macro-turbulence---and the result must be the ejection of a magnetic flux tube, called a plasmoid, from the region.

The example of a footpoint flow that was discussed here had a very simple and smooth spatial and temporal dependence; realistic footpoint motions of coronal loops are far more complicated, which makes them more prone to chaos.  However, the parts of the footpoint motion that cause the most rapid development of states in which large-scale reconnection is inevitable have a spatial scale comparable to the region $a$ over which reconnection will take place.
 
 The calculations made in this paper demonstrate that a magnetic field reaches a state in which reconnection is inevitable on a time scale that is only logarithmically longer than the ideal evolution time scale, even when drive for the evolution is simple.  
 
 Nonetheless, more complete simulations related to those of Reid et al \cite{Reid:2018}, are required for a more complete understanding of how the reconnection proceeds.  It is particularly important to use such simulations to understand how the magnetic reconnection region is determined, not because it is bounded by a rigid conductor but by unstressed magnetic field lines.  This can be done by choosing a non-zero $\lambda^2$ in Equation (\ref{h}).  As $\lambda^2$ is made larger within the model of this paper, simulations become more expensive but more representative of the corona.  Such simulations are numerically simpler when the circular cylinder of Figure \ref{fig:cylinder} is replaced by a cylinder of square cross section that encloses the volume with $|x|<a$, $|y|<a$, and $0<z<L$.  As mentioned at the end of Section \ref{sec:choice of v_t}, the only significant modification required is a change in the factors of $1-r^2/a^2$ in Equation (\ref{h}) to $(1-x^2/a^2)(1-y^2/a^2)$.

 
\section*{Acknowledgements}

 The authors wish to thank Yi-Min Huang for pointing out an error in the derivation of Equation (\ref{?K/?t}) in an earlier version of the manuscript.  This work was supported by the U.S. Department of Energy, Office of Science, Office of Fusion Energy Sciences under Award Numbers DE-FG02-95ER54333, DE-FG02-03ER54696, DE-SC0018424, and DE-SC0019479.

\section*{Data availability statement}

The data that support the findings of this study are available from the corresponding author upon reasonable request.



\end{document}